# Robust and high-resolution seismic complex trace analysis

M. Kazemnia Kakhki[a,b,*], W.J. Mansur[a,b], K. Aghazade[c]

[a] Modelling Methods in Engineering and Geophysics Laboratory (LAMEMO), COPPE, Federal university of Rio de Janeiro, 21941-596 RJ, Brazil

[b] Department of Civil Engineering, COPPE, Federal University of Rio de Janeiro, RJ, Brazil

[c] institute of geophysics, University of Tehran

**Abstract**

Seismic attributes calculated by conventional methods are susceptible to noise. Conventional filtering reduces the noise in the cost of losing the spectral bandwidth. The challenge of having a high-resolution and robust signal processing tool motivated us to propose a sparse time-frequency decomposition while is stabilized for random noise. The procedure initiates by using Sparsity-based adaptive S-transform to regularize abrupt variations in frequency content of the nonstationary signals. Then, considering the fact that a higher amplitude of a frequency component results in a higher signal to noise ratio, an adaptive filter is applied to the time-frequency spectrum which is sparcified previously. The proposed zero adaptive filter enhances the high amplitude frequency components while suppresses the lower ones. The performance of the proposed method is compared to the sparse S-transform and the robust window Hilbert transform in estimation of instantaneous attributes by applying on synthetic and real data sets. Seismic attributes estimated by the proposed method is superior to the conventional ones in terms of its robustness and high resolution image. The proposed approach has a vast application in interpretation and identification of geological structures.

## 1. Introduction

The data interpretation in signal analysis can be better accomplished if a different perspective of the data is available. This aim can be achieved by transforming the data from one domain to the other. The Fourier transform is one of those common transformations which enables us to survey average properties of a remarkably vast portion of a trace, although it does not represents local variation. The analytic signal, complex trace, first introduced in seismology by (Taner et al. 1979) and then developed by (Gabor 1947), resolved this problem by maintaining the local significance and also yields new perspective. The analytic signal is a complex signal whose imaginary part is the Hilbert transform of its real part (Gabor 1947).

Conventional seismic methods are incapable of deciphering subtle geological features, this fact have been instigated the researchers in excavation of various techniques to resolve this challenge. Instantaneous seismic attributes took advantage of complex trace to extend their definitions of simple harmonic oscillation and have been used in interpretation of structural features. Seismic attributes analysis are able to deal with stratigraphic and geological properties (Taner et al. 1979; Barnes 2007; Berthelot et al. 2013; Verma et al. 2018) since they provide quantitative measure of phase, frequency, and reflector amplitude such as the distribution of reef complexes which can be explained by instantaneous phase (Zheng et al. 2007). Thin-bed tuning is the other challenging structure which is possible to be detected by instantaneous frequency (Chopra and Marfurt 2005). The reservoir characterization and limestone formations were delineated via seismic instantaneous amplitude, frequency and phase by imaging various target units (Farfour et al. 2015). (Ali et al. 2019) used the dominant frequency attribute to define the characterization of hydrocarbon bearing reservoir.(Verma et al. 2018) inferred the dunal and interdunal deposits in 3D seismic data volume through the combination of coherence attribute and inverted P-impedance. Texture and edge attributes were used by (Asjad and Mohamed 2015) to extract salt dome.

The other significant stratigraphic exploration issue is the porous rocks bounded in a nonporous matrix. (Bedi and Toshniwal 2019) estimated the porosity of reservoir from seismic attributes that have good correlation with the property porosity (energy, Mean, Instantaneous Amplitude, Homogeneity, Autocorrelation, Cosine phase, Contrast, Dissimilarity and instantaneous frequency). (Takam Takougang et al. 2019) applied coherence and similarities attributes in delineating the fault and fractures from reverse time migrated seismic section. Channel boundaries are also detectable using seismic coherence and other edge sensitive attributes, although their thickness cannot be defined via these attributes. Hence spectral decomposition, which is sensitive to channel thickness, is used to complement the coherence and edge sensitive attributes (Partyka and Gridley 1999; Anees et al. 2019). (Obiadi et al. 2019) used spectral decomposition integrated with seismic attributes to identify the geometry and structural discontinuities of hydrocarbon reservoir within complex tectonic settings.

Although Seismic complex attributes are applicable in defining complex structures, they are problematic in noisy data due to their sensitivity to noise. To alleviate this defect, (Luo et al. 2003) presents a generalized version of Hilbert transform (HT). (Liu and Marfurt 2007) outlined the efficiency of time frequency representation (TFR) in achieving cleaner instantaneous frequency in thin-bed and channel detection. (Lu and Zhang 2013) introduced the windowed Hilbert transform (WHT), a time-frequency representation form of HT accompanied by a zero-phase adaptive filter to enhance instantaneous complex attributes. Despite the efficiency of filtering in removing the undesired frequency components in complex trace analysis, losing the original data is the primary concern. Concerning this fact (Sattari 2016) proposed a fast sparse S-transform to achieve sparse WHT by applying the optimized windows in frequency domain. Although the resolution of the seismic attributes has improved via sparse ST proposed by (Sattari 2016), the presence of random noise remains unsolved. Therefore, to achieve stable and high-resolution instantaneous spectral attributes, the fast sparse S transform (SST) was improved by the robust adaptive WHT (RAWHT) to achieve sparse RAWHT, which concerns the abrupt changes in the frequency content of the signal and is less sensitive to the noise. The role of the adaptive filter is to suppress the lower-amplitude frequency components and improve the higher-amplitudes.

In this contribution a modified calculation of analytic signal is presented to provide a robust Hilbert transform which is of high-resolution and indifferent to noise to have better estimation of instantaneous attributes. The main aim of the proposed method is using sparsity-based window-parameters optimization to improve the resolution of the seismic attributes while taking advantage of a zero phase adaptive filter to stable the contaminated data in calculating the analytic signal in time-frequency domain. The principal advantage of our method is to render higher resolution HT which is less sensitive to random noise.

In this work, we compared the performance of the proposed method, sparse WHT with the HT , the SST, and the RWHT on synthetic and real data. We begin with the explanation of complex trace analysis followed by the calculation of instantaneous attributes. Afterwards, analytic signal is improved by using sparse ST and zero phase adaptive filter and ending with the examples to compare the performance of the proposed method to the HT the SST, and RWHT.

## 2. Methodology

### 2.1 Calculation of the complex trace

Assume the real signal x(t) is x(t)=A(t) cosθ(t) and the imaginary part is y(t)=A(t) sinθ(t), the complex trace z(t) or the analytic signal is computed as

$$z(t) = x(t) + iy(t) = A(t)e^{j\theta(t)} \tag{1}$$

where y(t) is the HT of input signal x(t) derived from the convolution of x(t) with the function –(1/πt). Hilbert transform is considered as a linear, time-invariant system with impulse response

$$h(t) = -\left(\frac{1}{\pi t}\right) = \delta(t) \tag{2}$$

where δ(t) is the Dirac delta function. Signal x(t) can be analytic by applying the HT in the frequency domain. Considering X(w) as the Fourier spectrum of x(t) then Z(w), the spectrum of the analytic signal, is calculated as follows:

$$Z(\omega) = X(\omega)[1 + iH(\omega)] = \begin{cases} 2X(\omega) & \omega > 0 \\ X(\omega) & \omega = 0 \\ 0 & \omega < 0 \end{cases} \tag{3}$$

where H(ω) is the filter as

$$H(\omega) = \begin{cases} -i & \omega > 0 \\ 0 & \omega = 0 \\ i & \omega < 0 \end{cases} \tag{4}$$

The amplitude spectrum of the complex trace z(t) is double for positive frequencies, while it is zero for negative ones. Therefore, the complex trace can be formed by taking the Fourier transform of the real trace, make the amplitude of negative frequencies zero and double the amplitude of positive frequencies, and finally apply the inverse Fourier transform. Afterwards, the instantaneous frequency and phase are easily achievable via the analytic signal.

Seismic instantaneous attributes (Taner et al. 1979) can be derived from the analytic signal. A(t) and θ(t) denote the instantaneous amplitude and the instantaneous phase in equation 1, respectively. The instantaneous frequency can be obtain by taking the derivative of instantaneous phase

$$\varphi(t) = \frac{d\theta(t)}{dt} \tag{5}$$

The real seismic signal can have abrupt changes and interferences both in time and frequency because it carries the information of heterogeneous subsurface. Therefore it can affect the obtained analytic signal especially in terms of resolution. (Sattari 2016) attempt to address this problem via optimized windows and proposed sparse ST. (if you want to compare the HT with time-frequency methods put a figure here).

### 2.2 Analytic signal using sparse ST

We used sparse ST method proposed by (Sattari 2016) to calculate the analytic part of a signal to have higher resolution in comparison to the other known methods. The windowed HT can be defined in time-frequency domain as

$$Z(\omega,\tau) = X(\omega,\tau)[1 + iH(\omega)] = \begin{cases} 2X(\omega,\tau) & \omega > 0 \\ X(\omega,\tau) & \omega = 0 \\ 0 & \omega < 0 \end{cases} \quad (6)$$

The main difference between the adaptive sparse ST proposed by (Sattari 2016) with the standard ST proposed by (Stockwell et al. 1996) is that it uses frequency dependent window parameters reversely proportional to the amplitudes of various frequency components while the former uses window-length directly proportional to frequency while windowing the frequency domain input signal. The strategy used to obtain the adaptive sparse ST is that frequency components with higher amplitude are forced to dominate the time-frequency lattice by being localized while translation using high and short windows, whereas the lower amplitude harmonics need to be smeared in time-frequency domain by using low and wide windows while being translated. As a result, the computational complexity of the adaptive sparse ST and the standard ST is basically the same. Besides, the algorithm of the adaptive sparse ST has another notable difference in that it uses the above-mentioned window-parameters (height and length) optimization to create completely arbitrary windows to translate the spectrally varying signal with highest adaptivity ideally. (Sattari 2016) implemented this technique by first exploiting matrix formulation of ST as

$$\hat{x}[m] = \sum_{n=1}^{N} \sum_{k=1}^{N} TFR_{ST}^{[n]}[n,k][g^{s[m]}[m+n-1]\exp(-2i\pi mk/N)]_{m=0,\ldots,N-1} \quad (7)$$

where *g* is a window function shifted along the frequency axis by the step of the frequency shift m from 0 to *N*-1. *k* and *n* are the time and frequency indices, respectively, with the values, vary from 1 to *N*. *s[m]* is defined as the standard deviation of the window function in statistics which has the role of frequency dependent support of *g*.

Then taking advantage of the valuable information included in the input signal amplitude spectrum to distinguish between high and low amplitude frequency components according to their known positions, and finally by changing the optimization direction from frequency to frequency shift which enabled him to use the amplitude spectrum to create the above mentioned sparsity under the matrix formulation. According to the linear program provided by (Sattari 2016), the change in the optimization direction is performed by a simple transpose in the algorithm of the ST which results in standard voice Gaussians along frequency shift while smooth and differentiable arbitrary windows along frequency are automatically obtained. Fig.1 shows the performance of the Adaptive sparse ST along with the adaptive arbitrary windows used in that applied to a non-stationary signal compared to that of the ST.

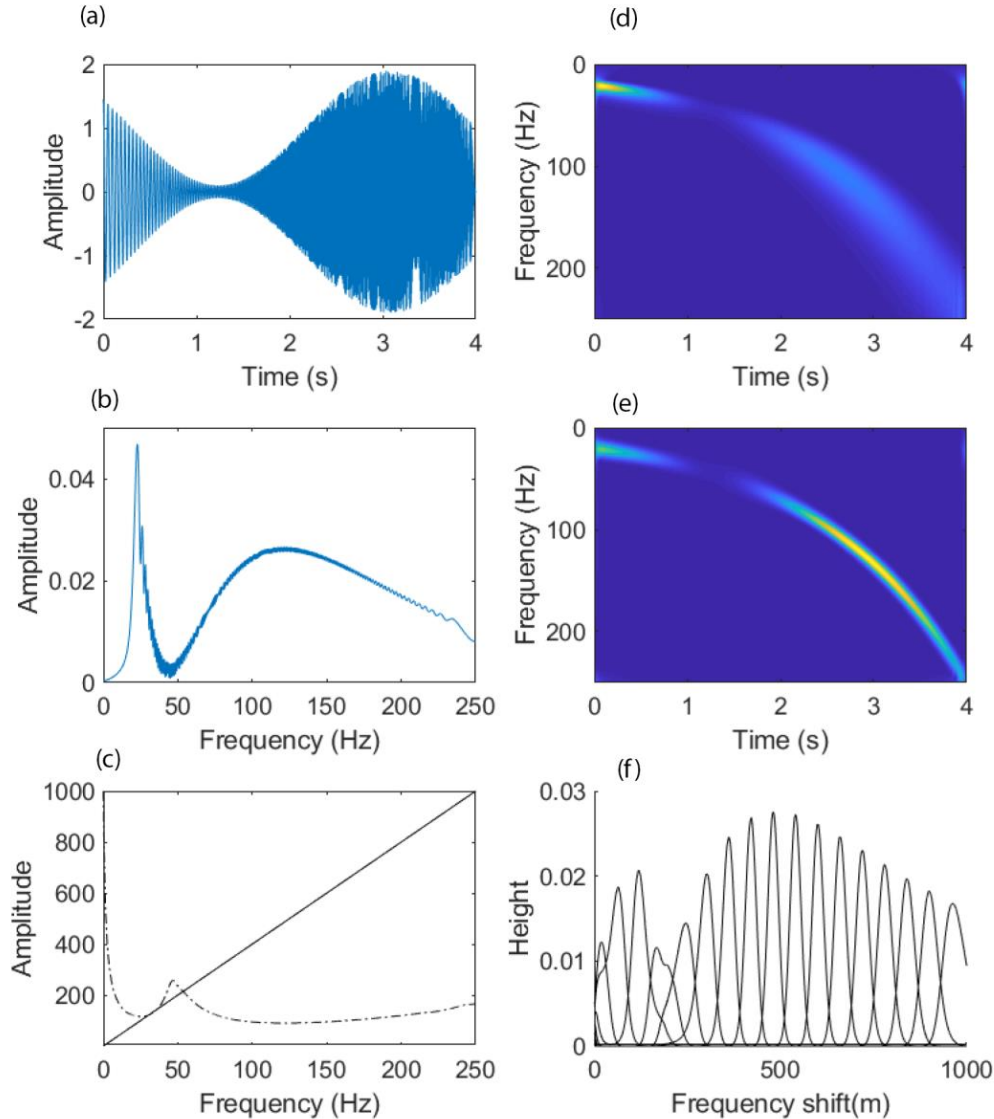

Figure 1 (a) A nonstationary logarithmic chirp signal with sinusoid variations in amplitude along time axis, (b) the correspondent amplitude spectrum, (c) the window length variation with frequency for standard ST (thick) and adaptive sparse ST (dashed). Time-frequency map of the non-stationary signal computed by (d) the conventional ST and (e) adaptive sparse. (f) The shifted adaptive arbitrary windows used in the adaptive sparse ST which are in high compatibility with the amplitude spectrum shown in subplot b while sliding over it for each shift

As a result, the adaptive sparse ST is not only superior to the standard ST in terms of adaptivity and higher resolution, but also it is very efficient in that it adds no extra computation to the translation and modulation processes required for the spectral decomposition. This means it even performs better than the alternative energy concentration (ECM) methods (Jones and Parks 1990; Sattari et al. 2013) used for adaptivity enhancement of Fourier-based spectral decomposition. These methods are computationally heavy as they require computation of several time-frequency decompositions with different window-lengths among which sparsest result is searched for while in the adaptive sparse ST, the window parameters are optimally set to create sparsity. This makes the ECM methods impractical for real-world application. In addition to the complexity, (Sattari 2016) also showed the superiority of the adaptive sparse ST over the standard ST and STFT optimized by ECM methods in terms of robustness to noise, temporal and spectral interference resolution and finally the fact that it has only one free parameter to set which is linear and well-behaved. However, under the low SNR the sparse ST is not stable. For these reasons, in this paper, the TFR obtained via adaptive sparse ST are filtered in the time-frequency domain.

2.3 Improved Hilbert transform

To resolve the problem of noise in the signal, a time-frequency adaptive filter is applied to the TFR obtained via adaptive sparse ST. This filter is based on the assumption that higher amplitude spectrum has more content of signal and is formed as

$$g(\omega, \tau) = \frac{|X(\omega, \tau)|^{N-1}}{\underset{\omega}{argmax}(|X(\omega, \tau)|^{N-1})} \quad (8)$$

where N is weighting factor, and N ≥ 1, $|X(\omega, \tau)|$ is the amplitude spectrum of X(ω,τ).

The analytic signal constructed as

$$z(t) = \int_{-\infty}^{\infty} F^{-1}(\tilde{Z}(\omega, \tau)) \, d\tau \quad (9)$$

Where

$$\tilde{Z}(\omega, \tau) = X(\omega, \tau)g(\omega, \tau)[1 + iH(\omega)] \quad (10)$$

$F^{-1}(\tilde{Z}(\omega, \tau))$ is the inverse Fourier transform of Z(ω,τ).

The increase in value of N results in amplification of frequencies with the maximum amplitude. The value of N depends on SNR, the higher the SNR, the lower the N. By applying N greater than one, the SST develops into the RSST with enhanced higher amplitude frequency components and suppressed lower ones. Although the SST is supposed to render less noisy results, it fails to suppress the noise when the SNR is low. On the other hand, the proposed adaptive filter by (Lu and Zhang 2013) cannot distinguish desired signal and undesired noise if applied directly to the TFR. The weight factor proposed by (Lu and Zhang 2013) reduces the noise at the cost of losing the signal and conclusively losing the subsurface information. Therefore applying a weighting order to the obtained TFR in SST not only can results high resolution TFR as SST but also can suppress the noise as well as RWHT with the difference that the signal is preserved.

The main application of the applied adaptive filter is enhancement of instantaneous phase estimation in noisy data, although it has application in improving the other seismic attributes. Figure 2 displays the failure by applying the SST, RAWHT to a synthetic noisy model.

3. Examples

The performance of the proposed method is validated by applying on synthetic and real data set. We compare our method in obtaining seismic attributes with the HT, SST, and RWHT method to observe the discrepancies in their performance.

Figure 1 shows the robustness of RSST in comparisons to sparse WHT and RWHT to a synthetic noisy wedge model.

Clearly, the adaptive sparse TFRs benefit from higher resolution and more adaptivity compared with sparse ST and RWHT.

### 3.1 Numerical examples

The choice of adaptive Fourier-based time-frequency decomposition is basically user dependent which is relevant to the type of analysis to focus in the time or frequency domain and more important than that is the characteristic of the signal (Radad et al. 2015). Seismic data as an example are narrow in frequency and wide in time, therefore, adaptive ST can be the best choice in analysis owing to the higher sparsity of the input frequency domain signal. Moreover, decomposing of the sparser version of the input signal can suppress the scattered random noise in both time and frequency domains more efficiently although not completely. These are the reasons of applying sparse ST in the first step to obtain high resolution spectral attributes. In the following step to stabilize our sparse TFR, a weight factor is added to suppress almost all the noise available in the TFR, according to our scope of analysis which results in a robust high resolution spectral amplitude.

To diagnose the superiority of the RSST over the SST and RWHT for decomposing narrowband signals, the TFR of 5 nonstationary signals are compared in figure 2. The signal taken from (Andrade et al. 2018) is a sum of five signals generated by a sampling interval of 0.003s as

$$\begin{cases} x_1(t) = 0.8\cos(30\pi t) & 0 \leq t \leq 6s \\ x_2(t) = 0.6\cos(70\pi t) & 0 \leq t \leq 6s \\ x_3(t) = 0.7\cos(130\pi t + 5\sin(2\pi t)) & 4s \leq t \leq 8s \\ x_4(t) = \sin\left(\dfrac{8\pi 100^{t/8}}{\log(100)}\right) & 6s < t \leq 10s \\ x_5(t) = 3e^{-1250(t-2)^2}\cos(710(t-2)) & 0 \leq t \leq 10s \end{cases} \quad (11)$$

Where $x_1$ is a harmonic component of 15 Hz, $x_2$ is another harmonic with 35 Hz, $x_3$ is a frequency-modulated harmonic around 65Hz, $x_4$ is a sliding harmonic from 35 to 158 Hz, and $x_5$ is a Morlet wavelet with central frequency of 113 Hz. The signal is shown in Figure 2a, its corresponding instantaneous frequencies in b, and their TFR obtained by SST, RWHT, and RSST, are in c to e, respectively.

As is seen, the TFRs obtained by RSST give spectra with higher resolution and more stable focused in time, frequency or both. It worth to mention that regularizing the true positions and amplitudes of different components in time-frequency domain results in more accurate instantaneous attributes since the scattered energy in the time-frequency domain can cause fake complex indices.

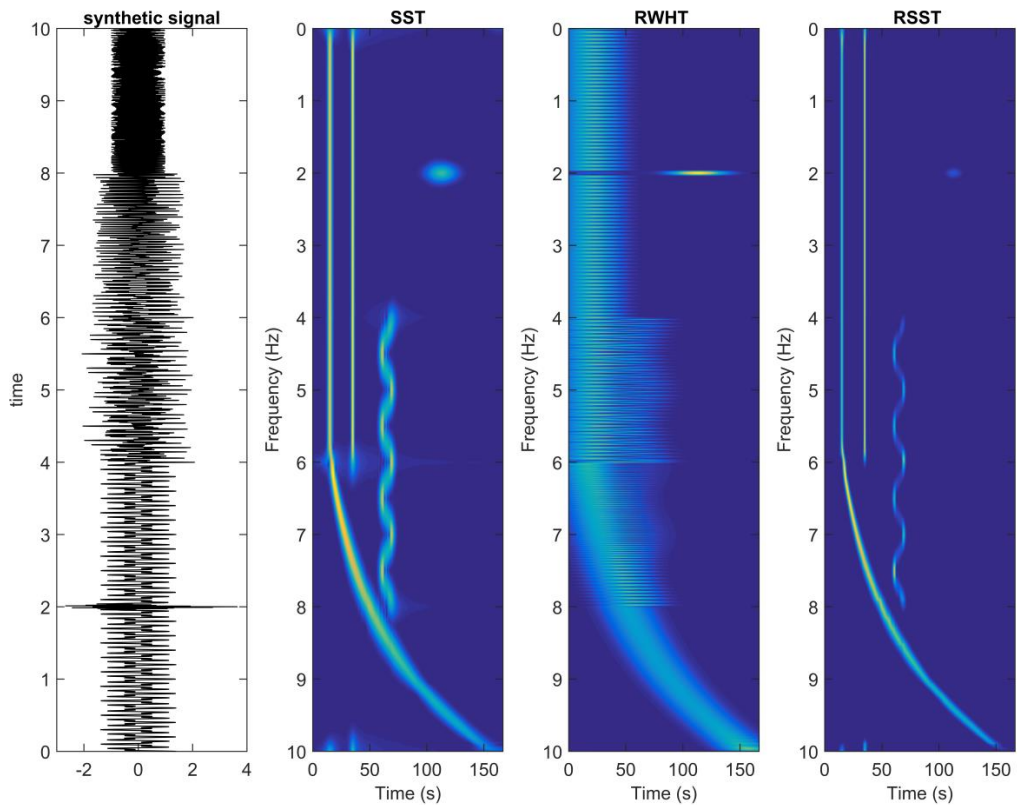

Figure 2. time-frequency distribution of (a) the synthetic seismogram (b) the model (c) SST (d) RWHT and (e) RSST. (khode signal ro ham bezar, age noise ezafe mikoni bezar, (b) ke gofte instantaneous frequency ke ye khate barike ro ham bezar)

Therefore, as seen in figure 2 the SST and RWHT are not as efficient as the proposed RSST is in sparsifying the TFRs and producing high resolution. The superiority of using sparse ST proposed by sattari to the conventional methods is that it reduces the effect of noisy components before optimizing window parameters during the smoothing process on the amplitude spectrum. The sparse base of this method in addition to its less susceptibility to random noise was the reasons of using it instead of conventional methods.

Although the proposed method by (Sattari 2016) reversely scales the global trend of the amplitude spectrum to be indifferent to the added noise, it is still susceptible to the random noise since the weight that the windows are applying to the signal is the same for the noises as well. On the other hand the other method is robust in detecting the noise, but in the cost of losing the signal. This point is depicted in figure 3 for a double chirp signal with additive random noise of 2 dB signal-to-noise ratio (in meghdaro chejuri hesab konim). This signal is chosen since it is wide in both domains to assure justified comparison between the proposed method SST, and RWHT. The parameters taken in the example are the same as for the three TFRs for smoothing (r=10) in SST and RSST and denoising (N=2.5) in RSST and RWHT.

(signal ro bezar, noise ezafe kon bezar, TFR ro vase raveshaye mokhtalef bezar, age lazem bud fft signal bedun noise va noise ro ham bezar)

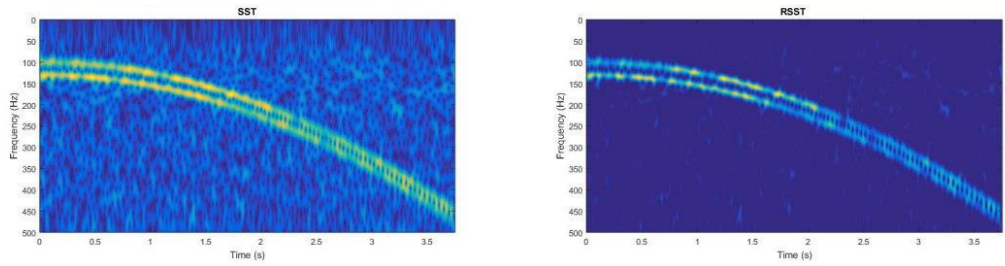

Figure 3 (a) Double chirp signal (b) its amplitude spectrum. (c) The same chirp signal with 2dB additive random noise (d) its amplitude spectrum. (e)-(g) the TFR of the clean signal obtained by SST, RWHT and RSST, respectively. (h)-(j) the TFR of the noisy signal obtained by SST, RWHT and RSST, respectively.

Considering the achievement of numerical examples, the RSST represents a higher-resolution and more robust to the noise compare to the SST and RWHT. The input signal in this method can be regularized in time-freqquency while estimating the analytic signal owing to its low susceptibility to random noise. The results of using the proposed method (RSST) to a synthetis wedge model are compared with those of SST, HT and RWHT in Figure 4.

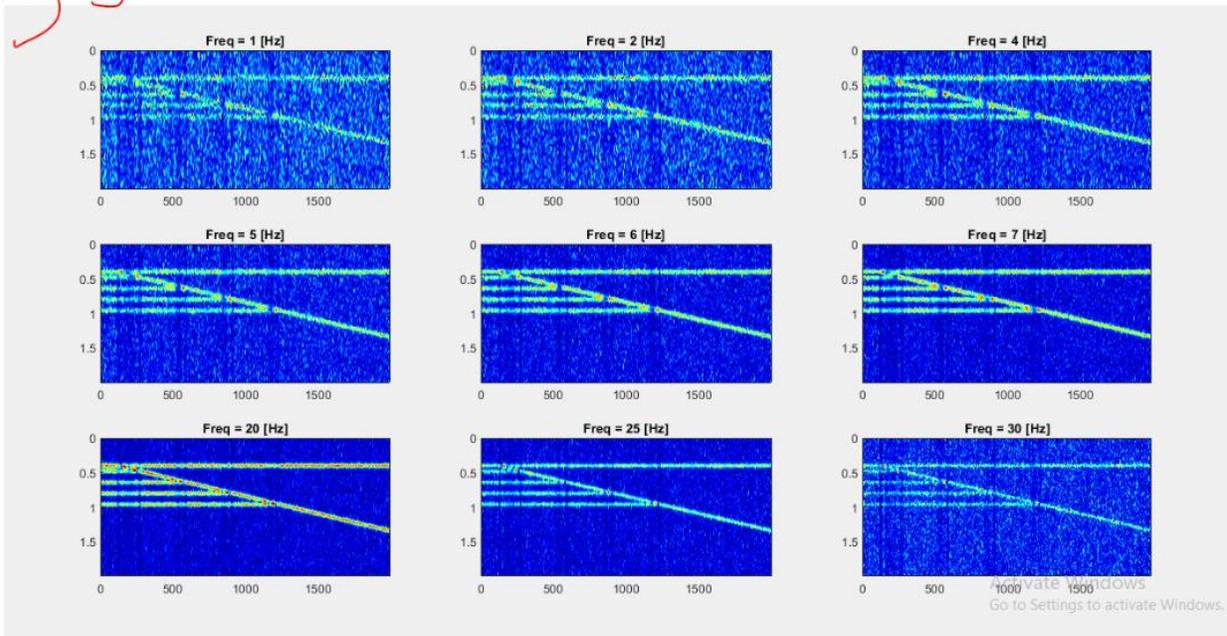

Figure 4. Complex trace analysis of a wedge model. (a) synthetic seismic section obtained by convolving the reflection coefficients with a 10 Hz Ricker wavelet, (b) the model with 2 dB additive random noise. (c) the instantaneous amplitude, instantaneous frequency, and cosine of instantaneous phase from left to right, respectively resulted by HT, (d) SST, (e) RWHT, and (f) RSST.

It can be seen that the layers of the noisy model achieved by the RSST is of higher resolution and is of lower noise. Moreover, the strength of RSST in detection of fake reflectors is obvious in cosine instantaneous phase section which is clear of these reflectors around the wedges (age vaghean dare neshun mide bezar jomle bashe). Obviously the resolution in RSST has improved via optimizing window parameters and rendering weight order to the signal in each selected window.

### 3.2 Real example

The proposed method is applied to real data including ( what is inside the real data to detect). The first example displayed in Figure 5 is excerpted from a 2D seismic data acquired in … with known ( gas reservoirs embedded in thin-bed layers of an anticline.

Figures 14,15 present the results of complex trace analysis via different methods.( age khasti har attribute rot u ye figure bezari, v aba HT moghayese koni)

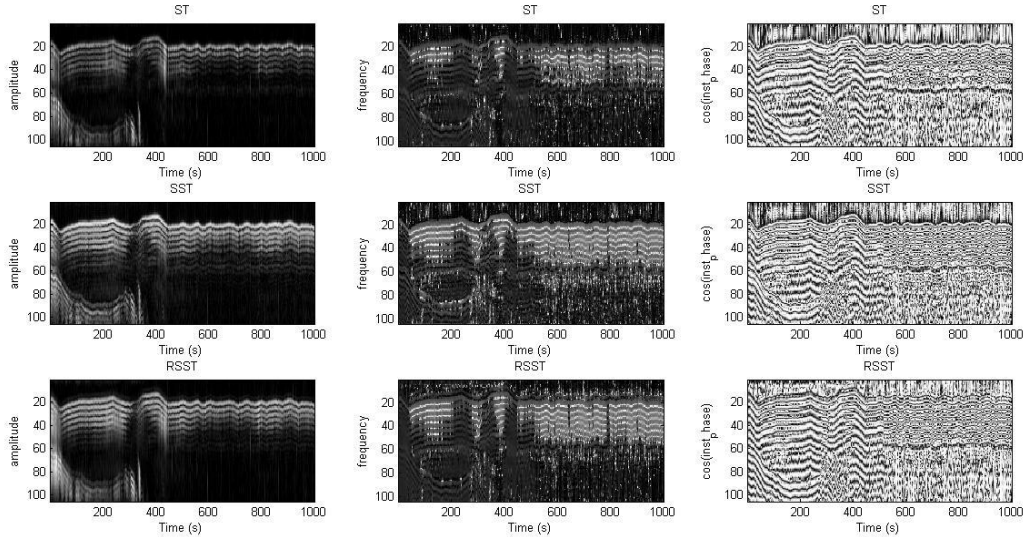

The known (reservoirs ) are detected in Figure 13b by the means of low-frequency shadow (Ebrom, 2004). Consider that the high amplitude anomalies related to (gas reservoir) delineated in the 10 Hz monofrequency section disappear in higher frequency monofrequency sections (if monofrequency is useful).

Obviously, the instantaneous attributes estimated by the RSST are of the highest quality and resolution in comparison with the ones obtained by other methods. The amplitude envelope is used to detect relevant seismic units via their energy content. As for the instantaneous amplitude achieved by the RSST in Figure 14d, (the bright spot at the apex of the anticline) that is resulted from one of the (gas reservoirs) has been detected with higher resolution and more stable compared with those obtained by the other methods.

The seismic energy absorption feasibly relevant to the ( presence of gas in the regions) compatible to the positions of the known (gas reservoirs) results in an abrupt reduction in the instantaneous frequency(Figure 15). The dark anomalies illustrated in Figure 15d (instantaneous frequency section) achieved by the RSST are related to the stratigraphic boundaries of (the reservoirs).

Moreover, the strength of the RSST in delineating the lateral changes of the reflectors by cosine of instantaneous phase is of significant interpretational aims. To define the efficiency of the proposed method in regularizing the signal in time frequency domain, TFRs obtained by different methods of trace number (70) of the real data as an input signal are compared in Figure 6. As can be seen in Figure 6b, the nonfiltered WHT (SST) ignores the presence of noise and abrupt changes in frequency content, while filtering via RWHT smoothens them. The filtering applied in RWHT enhances the peak frequency contribution to the reconstructed analytic signal while causes the loss of the initial spectral bandwidth. Of the three methods compared, the TFR obtained by the RSST (Figure 6d) is more robust and efficient in resolving interfered wavelets under the random noise while maintaining the preliminary bandwidth. The cosine of instantaneous phase achieved by the RSST proves the power of proposed method in resolving interfered wavelets with details.

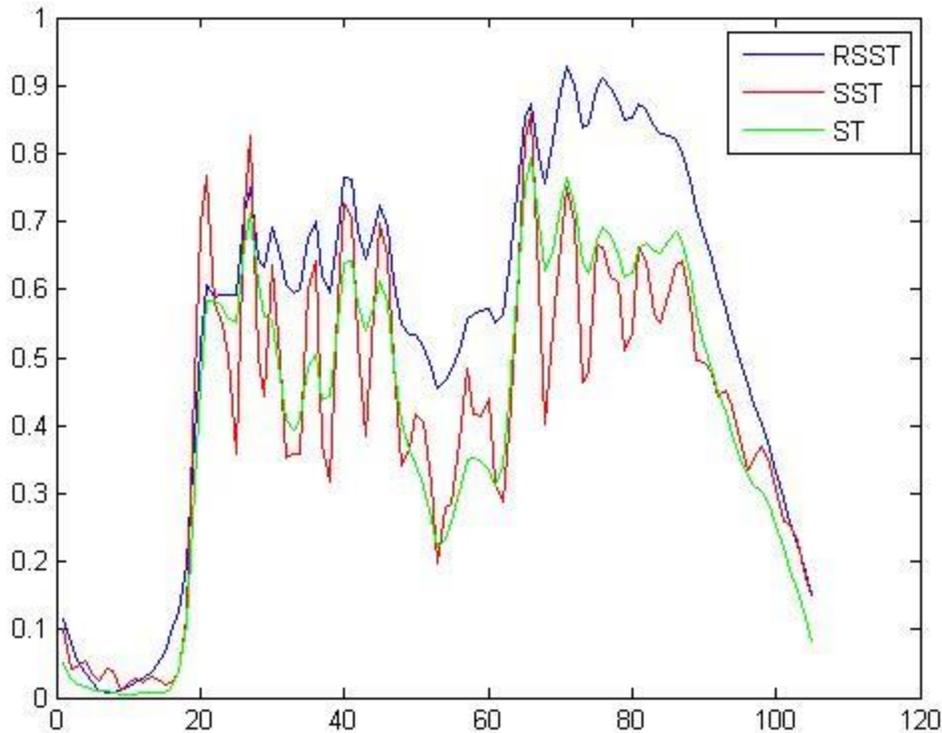

## 4. Conclusion

In this paper, the problem of estimation stable and high-resolution complex trace analysis was studied in the framework of sparsity-based optimization and time-frequency spectrum weighting order. Considering the time-frequency domain, the resolution and noise problem shown the necessity of a more strength method to deal with them. Therefore, the Sparsity-based adaptive S-transform was proposed as a spectral decomposition tool to enhance the resolution of the time-frequency WHT. The optimized windows satisfy the requirements for regularization of abrupt frequency changes and have superiority to the previous methods in terms of computation cost and interference removal without leading to fake indices. The proposed spectral decomposition is then enhanced via a zero phase adaptive filter to suppress the residue noise failed to prevent by the spectral decomposition method by enhancing the frequency components with larger amplitudes. As for the computational cost, the proposed method is slightly slower than SST because of an extra denoising in TFR; however, it is faster than RWHT due to an additional inverse Fourier transform of it.

Conclusively the proposed robust spectral decomposition approach was used to implement complex trace analysis of synthetic and real data sets. The results proved that the power of robust adaptive ST in regularizing the abrupt frequency changes and suppression of random noise results in high resolution and robust instantaneous attributes compare to the conventional methods that ignore these changes. Indeed, the proposed method regularizes the entire frequency content of the signal by setting only one window parameter and suppresses the noise spread in both time and frequency domain with adjusting the weighting order N. It should be mentioned that the proposed method as an adaptive, high resolution, invertible and frequency-dependent time-frequency decomposition approach has a vast application in interpretation and complex trace analysis.it should be mentioned that the reason of using ST in this procedure as a spectral decomposition is relevant to the input data and corresponding application and other transforms like STFT is also possible to be used.